\documentclass[preprintnumbers,amsmath,amssymbm,prd]{revtex4}
\usepackage{epsfig}
\usepackage{graphicx}

\begin{document}
\title{Long-lived resonances of massive scalar fields in the Reissner-Nordstr\"om black-hole spacetime: Analytic treatment 
in the large-mass regime}
\author{Shahar Hod}
\affiliation{The Ruppin Academic Center, Emeq Hefer 40250, Israel}
\affiliation{ }
\affiliation{The Jerusalem Multidisciplinary Institute, Jerusalem 91010, Israel}
\date{\today}

\begin{abstract}
\ \ \ The physical and mathematical properties of the composed 
Reissner-Nordstr\"om-black-hole-massive-scalar-field system are studied {\it analytically} in the dimensionless 
large-mass $M\mu\gg1$ regime 
[here $\{M,\mu\}$ are respectively the mass of the central black hole and the proper mass of the scalar field]. 
It is proved that, for a given value ${\bar Q}\equiv Q/M$ of the dimensionless charge parameter 
of the central black hole, the system is characterized by the presence of quasi-resonances, 
linearized perturbation modes with arbitrarily long lifetimes. 
In particular, using analytical techniques, 
we determine the black-hole-field critical mass spectrum $\{M\mu_{\text{crit}}({\bar Q})\}$ 
which characterizes the long-lived resonances of the composed physical system.
\end{abstract}
\bigskip
\maketitle

\section{Introduction}

The dynamics of matter and radiation fields in curved black-hole spacetimes are usually characterized 
by damped quasinormal resonant modes whose 
complex eigen-frequencies $\{\omega_n\}^{n=\infty}_{n=0}$ may provide 
physically valuable information about the properties of the central black hole. 
These unique resonances are of particular importance in observational studies as well as in theoretical explorations 
of the physical properties of composed 
black-hole-field systems 
(see \cite{Nollert1,Press,Cruz,Vish,Davis,Mash,LV,QNMs,Detwe,Onoz} and references therein). 

The temporal decay of generic black-hole-field resonant modes reflect the fact that most fundamental fields 
(and, in particular, fields that satisfy various no-hair theorems \cite{Whee,Car,Bek1,Chas,BekVec,Hart,Hod11}) 
cannot form spatially regular static matter configurations outside spatially regular horizons 
of asymptotically flat black holes \cite{Notescc,Hodrc,HerR}. 
In particular, the exponentially decaying quasinormal oscillation modes describe radiation and matter 
fields whose amplitudes decay in time, $\psi\sim e^{-\Im\omega\cdot t}$ \cite{Noteiw,Notetails,Tails}, 
as they radiate their energies to asymptotically far regions or into the central absorbing black holes \cite{Detwe}. 

The asymptotically measured timescale $\tau_{\text{relax}}$ that 
characterizes the relaxation dynamics of a 
perturbed black-hole-field system is determined by the mathematically compact relation \cite{Noteunit}
\begin{equation}\label{Eq1}
\tau_{\text{relax}}\equiv {{1}\over{\Im\omega_0}}\  .
\end{equation}
Here $\Im\omega_0$, whose value depends on the physical parameters of 
the central black hole and the perturbation field, is the imaginary part of the fundamental ($n=0$) 
black-hole-field resonant frequency. 

Intriguingly, the physically important studies \cite{ELL1,ELL2} have explicitly demonstrated numerically 
that composed black-hole-massive-scalar-field systems are characterized by a unique family of 
extremely long-lived $(\tau_{\text{relax}}\to\infty)$ quasi-resonant modes with the property   
\begin{equation}\label{Eq2}
\Im\omega_0(M\mu,l)\to0\ \ \ \ \text{for}\ \ \ \ M\mu\to M\mu_{\text{crit}}(l)\  .
\end{equation}
Here $M$ is the mass of the 
central black hole and $\{\mu,l\}$ are respectively the proper mass \cite{Notemm} 
and the angular harmonic index [see Eq. (\ref{Eq8}) below] that characterize the linearized field mode. 

To the best of our knowledge, in the physics literature there are no closed form {\it analytical} formulas that describe 
the dimensionless critical mass spectrum $\{M\mu_{\text{crit}}\}$ which characterizes the long-lived quasi-resonant 
modes (\ref{Eq2}) of composed black-hole-massive-field systems.

The main goal of the present compact paper is to determine the parameter-dependent 
timescales that characterize the relaxation dynamics of minimally coupled massive scalar 
fields in the asymptotically flat Reissner-Nordstr\"om black-hole
spacetime \cite{Chan}. 
Interestingly, as we shall explicitly prove below, the composed 
Reissner-Nordstr\"om-black-hole-massive-scalar-field system is amenable to an analytical treatment 
in the dimensionless regime  
\begin{equation}\label{Eq3}
M\mu\gg1\ \ \ \ \text{with}\ \ \ \ l\gg1\
\end{equation}
of large black-hole-field masses and large angular indices.

In particular, below we shall explicitly prove that the resonance spectrum $\{M\mu_{\text{crit}}(Q/M,l)\}$ 
of critical black-hole-field masses, which characterizes the long-lived modes (\ref{Eq2}) 
of the composed Reissner-Nordstr\"om-black-hole-massive-scalar-field system, 
can be determined {\it analytically} in the double asymptomatic regime (\ref{Eq3}).  

\section{Description of the system}

We shall analyze the physical and mathematical properties of a linearized massive scalar field which is minimally coupled 
to a curved Reissner-Nordstr\"om black-hole spacetime. 
The curved spacetime is characterized by the spherically symmetric line element \cite{Chan}
\begin{equation}\label{Eq4}
ds^2=-f(r)dt^2+{1\over{f(r)}}dr^2+r^2(d\theta^2+\sin^2\theta
d\phi^2)\  .
\end{equation}
The dimensionless metric function in (\ref{Eq4}) is given by the radially-dependent 
functional expression
\begin{equation}\label{Eq5}
f(r)=1-{{2M}\over{r}}+{{Q^2}\over{r^2}}\  ,
\end{equation}
where $Q$ is the electric charge of the central black hole \cite{Notegen}. 
The extremal Reissner-Nordstr\"om spacetime is characterized by the charge-mass relation $Q=M$. 
The horizon radii of the central black hole, 
\begin{equation}\label{Eq6}
r_{\pm}=M\pm(M^2-Q^2)^{1/2}\  ,
\end{equation}
are determined by the roots of the metric function $f(r)$.

The dynamics of the massive scalar field $\Psi$ in the curved black-hole spacetime (\ref{Eq4}) is 
governed by the Klein-Gordon wave equation
\cite{Stro,HodPirpam}
\begin{equation}\label{Eq7}
(\nabla^\nu\nabla_{\nu}-\mu^2)\Psi=0\  .
\end{equation}
Using the scalar field decomposition
\cite{Noteom}
\begin{equation}\label{Eq8}
\Psi(t,r,\theta,\phi)=\int\sum_{lm}e^{im\phi}S_{lm}(\theta)R_{lm}(r;\omega)e^{-i\omega
t} d\omega\ ,
\end{equation}
one obtains from Eqs. (\ref{Eq4}), (\ref{Eq5}), and (\ref{Eq7}) 
the ordinary differential equation 
\begin{equation}\label{Eq9}
\Delta{{d} \over{dr}}\Big(\Delta{{dR}\over{dr}}\Big)+UR=0\
\end{equation}
for the radial part $R(r)$ of the massive scalar eigenfunction. 
Here
\begin{equation}\label{Eq10}
\Delta=r^2f(r)\
\end{equation}
and
\begin{equation}\label{Eq11}
U=\omega^2 r^4 -\Delta(\mu^2r^2+K_l)\  ,
\end{equation}
where $K_l=l(l+1)$ (with $l\geq |m|$) is the familiar angular 
eigenvalue of the angular scalar eigenfunction $S_{lm}(\theta)$ \cite{Heun,Abram,Notesy}.

The radial differential equation (\ref{Eq9}) can be written in the form
\begin{equation}\label{Eq12}
{{d^2\psi}\over{dy^2}}-(V-\omega^2)\psi=0\  ,
\end{equation}
where 
\begin{equation}\label{Eq13}
\psi=rR\
\end{equation}
and the ``tortoise" radial coordinate $y$ is defined by the differential
relation \cite{Norery}
\begin{equation}\label{Eq14}
dy={{dr}\over{f(r)}}\  .
\end{equation}
The composed black-hole-massive-scalar-field radial potential in the Schr\"odinger-like ordinary differential 
equation (\ref{Eq12}) is given by the radially-dependent functional expression
\begin{equation}\label{Eq15}
V=V(r;M,Q,\mu,l)=f(r)\Big[\mu^2+{{l(l+1)}\over{r^2}}+{{2M}\over{r^3}}-{{2Q^2}\over{r^4}}\Big]\  .
\end{equation}

The discrete quasinormal resonant spectrum 
$\{\omega_n(M,Q,\mu,l)\}_{n=0}^{n=\infty}$, which characterizes the 
relaxation dynamics of the linearized massive scalar field in the curved black-hole spacetime (\ref{Eq4}), is 
determined by the Schr\"odinger-like differential equation (\ref{Eq12}) with the two 
boundary conditions 
\begin{equation}\label{Eq16}
\psi \sim e^{-i \omega y}\ \ \ \ \text{for}\ \ \ \ r\rightarrow r_+\ \ (y\rightarrow -\infty)\
\end{equation}
and
\begin{equation}\label{Eq17}
\psi \sim e^{i\sqrt{\omega^2-\mu^2} y}\ \ \ \ \text{for}\ \ \ \ r\rightarrow\infty\ \ (y\rightarrow \infty)\  .
\end{equation}
Note that the boundary condition (\ref{Eq16}) describes purely ingoing waves at the outer horizon 
of the central absorbing black hole, whereas the boundary condition (\ref{Eq17}) 
describes purely outgoing waves at spatial infinity for 
$\omega^2>\mu^2$ and asymptotically bounded (finite) scalar eigenfunctions for $\omega^2<\mu^2$ 
\cite{Detwe}.

In the next sections we shall study, using analytical techniques, the 
physical and mathematical properties of the composed 
Reissner-Nordstr\"om-black-hole-massive-scalar-field system in the dimensionless regime [see Eq. (\ref{Eq3})]
\begin{equation}\label{Eq18}
M\mu\gg1\ \ \ \ \text{with}\ \ \ \ l\gg1\
\end{equation}
of large black-hole-field masses and large angular indices. 
In particular, below we shall explicitly prove that the dimensionless 
spectrum $\{M\mu_{\text{crit}}(\Im\omega\to0)\} $ of critical 
black-hole-field masses, which characterizes the 
long-lived quasi-resonant modes of the composed physical system, can be determined {\it analytically} in the 
double asymptotic regime (\ref{Eq18}). 

\section{Black-hole-field resonance spectra: Analytic treatments for 
Schwarzschild and extremal Reissner-Nordstr\"om 
black holes} 

In the present section we shall explicitly show that the complex resonant frequencies 
that characterize the relaxation dynamics of 
massive scalar fields in the neutral Schwarzschild spacetime and in the maximally-charged (extremal) 
Reissner-Nordstr\"om spacetime can be determined {\it analytically} in the double asymptotic regime (\ref{Eq18}) 
of large black-hole-field masses and large angular indices. 

In the dimensionless large-mass-large-angular-momentum regime (\ref{Eq18}), the 
effective radial potential (\ref{Eq15}) of the composed black-hole-massive-field system can be approximated by
\begin{equation}\label{Eq19}
V=V(r;M,Q,\mu,l)=\Big(1-{{2M}\over{r}}+{{Q^2}\over{r^2}}\Big)\Big[\mu^2+{{l(l+1)}\over{r^2}}\Big]\cdot 
\big\{1+O[(M\mu)^{-2},l^{-2}]\big\}\  .
\end{equation}

As nicely shown in \cite{WKB1,WKB2,WKB3,Will}, the fundamental complex resonant frequencies 
of the Schr\"odinger-like radial differential equation (\ref{Eq12}) are determined, 
in the eikonal large-frequency regime, by the discrete resonance formula 
\begin{equation}\label{Eq20}
\omega^2=V_0-i(n+{1\over 2})\cdot\sqrt{-2V^{(2)}_0}\ \ \ \ ; \ \ \ \ n=0,1,2,...\  ,
\end{equation}
where the various derivatives $V^{(k)}_0\equiv d^{k}V/dy^{k}$ (with
$k\geq0$) that appear in (\ref{Eq20}) are evaluated at the radial peak $y=y_0(r_0)$,
\begin{equation}\label{Eq21}
{{dV}\over{dy}}=0\ \ \ \ \text{for}\ \ \ \ y=y_0\  ,
\end{equation}
of the effective black-hole-field potential. 

Using the frequency relation $\omega=\omega_{\text{R}}-i\omega_{\text{I}}$ and assuming 
the strong inequality [see Eqs. (\ref{Eq29}), (\ref{Eq30}), (\ref{Eq36}), and (\ref{Eq37}) below]
\begin{equation}\label{Eq22}
\omega_{\text{I}}\ll\omega_{\text{R}}\
\end{equation}
in the eikonal regime (\ref{Eq18}), one may decouple the real and imaginary parts of the resonance 
equation (\ref{Eq20}). In particular, one finds from the WKB condition (\ref{Eq20}) the functional relations
\begin{equation}\label{Eq23}
\omega_{\text{R}}=\sqrt{V_0}\cdot\{1+O[(M\mu)^{-1}]\}\
\end{equation}
and
\begin{equation}\label{Eq24}
\omega_{\text{I}}=\sqrt{{{-V^{(2)}_0}\over{2V_0}}}\cdot(n+{1\over 2})\cdot\{1+O[(M\mu)^{-1}]\}\ 
\end{equation}
for the real and imaginary parts of the black-hole-field resonant frequencies in the large-mass (large-frequency) 
regime (\ref{Eq18}). 

We shall now derive remarkably compact analytical formulas for the 
complex resonant spectra that characterize, in the dimensionless large-mass 
regime (\ref{Eq18}), the relaxation dynamics of 
massive scalar fields in the two opposite regimes: (1) The neutral ($Q\to0$) limit 
of Schwarzschild black-hole spacetimes, and 
(2) The maximally-charged ($Q\to M$) limit of extremal Reissner-Nordstr\"om black-hole spacetimes.  

\subsection{Resonance spectrum of the composed Schwarzschild-black-hole-massive-scalar-field system 
in the $M\mu,l\gg1$ regime}

In the present subsection we shall analyze the resonant spectrum of the composed 
Schwarzschild-black-hole-massive-scalar-field system, 
in which case the black-hole-field potential is given by the radially-dependent functional expression [see Eq. (\ref{Eq19})]
\begin{equation}\label{Eq25}
V=V(r;M,Q=0,\mu,l)=\Big(1-{{2M}\over{r}}\Big)\Big[\mu^2+{{l(l+1)}\over{r^2}}\Big]\  .
\end{equation}

Substituting Eq. (\ref{Eq25}) into the gradient relation (\ref{Eq21}) and using Eq. (\ref{Eq14}), 
one finds the two solutions
\begin{equation}\label{Eq26}
{\bar r}^{\pm}_0={{1\pm\sqrt{1-12{\bar\mu}^2}}\over{2{\bar\mu}^2}}\  ,
\end{equation}
where we have used here the dimensionless physical variables
\begin{equation}\label{Eq27}
{\bar\mu}\equiv {{M\mu}\over{\sqrt{l(l+1)}}}\
\end{equation}
and
\begin{equation}\label{Eq28}
{\bar r}\equiv {{r}\over{M}}\  .
\end{equation}
The inner radius in (\ref{Eq27}) corresponds to the local maximum point 
of the effective radial potential (\ref{Eq25}). 

Taking cognizance of Eqs. (\ref{Eq23}), (\ref{Eq24}), (\ref{Eq25}), and (\ref{Eq26}), 
one obtains the dimensionless 
functional relations \cite{Notesd}
\begin{equation}\label{Eq29}
{\bar\omega}_{\text{R}}({\bar\mu})=M\mu\cdot{{\sqrt{2}\big(1-\sqrt{1-12{\bar\mu}^2}-4{\bar\mu}^2\big)}\over
{\big(1-\sqrt{1-12{\bar\mu}^2}\big)^{3/2}}}\
\end{equation}
and
\begin{equation}\label{Eq30}
{\bar\omega}_{\text{I}}({\bar\mu})={{2\sqrt{2}{\bar\mu}^3
\sqrt{\sqrt{1-12{\bar\mu}^2}-(1-12{\bar\mu}^2)}}\over
{\big(1-\sqrt{1-12{\bar\mu}^2}\big)^2}}\cdot(n+{1\over 2})\
\end{equation}
for the real and imaginary parts of the black-hole-field resonant frequencies, where 
\begin{equation}\label{Eq31}
{\bar\omega}\equiv M\omega\  .
\end{equation}

From the analytically derived relation (\ref{Eq30}) one finds that ${\bar\omega}_{\text{I}}({\bar\mu})$ is a 
monotonically decreasing function of the composed black-hole-field mass parameter:
\begin{equation}\label{Eq32}
{{d{\bar\omega}_{\text{I}}}\over{d{\bar\mu}}}<0\ \ \ \ \text{for}\ \ \ \ {\bar\mu}<{{1}\over{\sqrt{12}}}\  .
\end{equation}
In particular, one deduces from (\ref{Eq30}) that 
the composed Schwarzschild-black-hole-massive-scalar-field system is characterized by the existence 
of arbitrarily long-lived quasi-resonant modes with \cite{Notetr1}
\begin{equation}\label{Eq33}
{\bar\omega}_{\text{I}}\to0\ \ \ \ \text{for}\ \ \ \ {\bar\mu}\to{\bar\mu}^{\text{Sch}}
_{\text{c}}={{1}\over{\sqrt{12}}}\  .
\end{equation}

\subsection{Resonance spectrum of the composed extremal-Reissner-Nordstr\"om-black-hole-massive-scalar-field 
system in the $M\mu,l\gg1$ regime}

In the present subsection we shall determine analytically the complex resonant spectrum of the composed 
extremal-Reissner-Nordstr\"om-black-hole-massive-scalar-field system, 
in which case the effective black-hole-field radial potential is given by the functional expression 
[see Eq. (\ref{Eq19}) with $Q=M$]
\begin{equation}\label{Eq34}
V=V(r;M,Q=M,\mu,l)=\Big(1-{{M}\over{r}}\Big)^2\Big[\mu^2+{{l(l+1)}\over{r^2}}\Big]\  .
\end{equation}

Substituting Eq. (\ref{Eq34}) into Eq. (\ref{Eq21}) and using the differential 
relation (\ref{Eq14}), one finds the two dimensionless radial solutions
\begin{equation}\label{Eq35}
{\bar r}^{\pm}_0={{1\pm\sqrt{1-8{\bar\mu}^2}}\over{2{\bar\mu}^2}}\  ,
\end{equation}
where the smaller radius in (\ref{Eq35}) corresponds to the local maximum point 
of the effective black-hole-field potential (\ref{Eq34}). 

Taking cognizance of Eqs. (\ref{Eq23}), (\ref{Eq24}), (\ref{Eq34}), and (\ref{Eq35}), 
one obtains the dimensionless functional relations \cite{Notesd}
\begin{equation}\label{Eq36}
{\bar\omega}_{\text{R}}({\bar\mu})=M\mu\cdot{{\sqrt{2}
\big(1-\sqrt{1-8{\bar\mu}^2}-2{\bar\mu}^2\big)^{3/2}}\over{\big(1-\sqrt{1-8{\bar\mu}^2}\big)^{2}}}\
\end{equation}
and
\begin{equation}\label{Eq37}
{\bar\omega}_{\text{I}}({\bar\mu})={{4{\bar\mu}^3\sqrt{[1-9{\bar\mu}^2+8{\bar\mu}^4-(1-5{\bar\mu}^2)
\sqrt{1-8{\bar\mu}^2}][1-\sqrt{1-8{\bar\mu}^2}-2{\bar\mu}^2]}}\over
{\big(1-\sqrt{1-8{\bar\mu}^2}\big)^3}}\cdot(n+{1\over 2})\  .
\end{equation}

From Eq. (\ref{Eq37}) one finds that ${\bar\omega}_{\text{I}}({\bar\mu})$ is a 
monotonically decreasing function of the dimensionless black-hole-field mass parameter:
\begin{equation}\label{Eq38}
{{d{\bar\omega}_{\text{I}}}\over{d{\bar\mu}}}<0\ \ \ \ \text{for}\ \ \ \ {\bar\mu}<{{1}\over{\sqrt{8}}}\  .
\end{equation}
In particular, from (\ref{Eq37}) one deduces that 
the composed extremal-Reissner-Nordstr\"om-black-hole-massive-scalar-field system is characterized 
by the existence of arbitrarily long-lived resonances with \cite{Notetr2} 
\begin{equation}\label{Eq39}
{\bar\omega}_{\text{I}}\to0\ \ \ \ \text{for}\ \ \ \ {\bar\mu}\to{\bar\mu}^{\text{eRN}}_{\text{c}}=
{{1}\over{\sqrt{8}}}\  .
\end{equation}

\section{Long-lived resonances of composed Reissner-Nordstr\"om-black-hole-massive-scalar-field systems 
in the general $Q/M\in[0,1]$ case}

In the present section we shall analyze the physical and mathematical properties of the composed 
Reissner-Nordstr\"om-black-hole-massive-scalar-field system for general values, 
\begin{equation}\label{Eq40}
{\bar Q}\equiv {{Q}\over{M}}\in[0,1]\  ,
\end{equation}
of the black-hole dimensionless charge parameter. 
In particular, our main goal in the present section is to determine, using analytical techniques, 
the charge-dependent critical mass spectrum $\{{\bar\mu}^{\text{RN}}_{\text{crit}}({\bar Q})\}$ 
which, in the dimensionless large-mass regime (\ref{Eq18}), 
characterizes the long-lived quasi-resonant modes of the composed black-hole-field system. 

Taking cognizance of Eqs. (\ref{Eq2}), (\ref{Eq21}), and (\ref{Eq24}) one deduces that the critical black-hole-field mass 
parameter ${\bar\mu}^{\text{RN}}_{\text{crit}}({\bar Q})$ of the quasi-resonant modes is 
characterized by the degenerate functional relations
\begin{equation}\label{Eq41}
\Big\{{{dV(y;{\bar Q},l,{\bar\mu})}\over{dy}}=0
\ \ \ \ \text{with}\ \ \ \ {{d^2V(y;{\bar Q},l,{\bar\mu})}\over{dy^2}}=0\Big\}
\ \ \ \ \ \text{for}\ \ \ \ \ {\bar\mu}={\bar\mu}^{\text{RN}}_{\text{crit}}({\bar Q})\  .
\end{equation}
Substituting the effective radial potential (\ref{Eq19}) into Eq. (\ref{Eq41}) and using the 
differential relation (\ref{Eq14}), one obtains the two coupled cubic polynomial equations
\begin{equation}\label{Eq42}
{\bar\mu}^2\cdot {\bar r}^3-(1+{\bar\mu}^2{\bar Q}^2)\cdot {\bar r}^2+3\cdot {\bar r}-2{\bar Q}^2=0\
\end{equation}
and
\begin{equation}\label{Eq43}
{\bar\mu}^2\cdot {\bar r}^3-{{3}\over{2}}(1+{\bar\mu}^2{\bar Q}^2)\cdot {\bar r}^2+
6\cdot {\bar r}-5{\bar Q}^2=0\
\end{equation}
for the dimensionless variables ${\bar r}={\bar r}_{\text{crit}}({\bar Q})$ 
and ${\bar\mu}={\bar\mu}_{\text{crit}}({\bar Q})$. 

From Eq. (\ref{Eq42}) one finds the charge-dependent relation
\begin{equation}\label{Eq44}
{\bar\mu}_{\text{crit}}=\sqrt{{{{\bar r}^2_{\text{crit}}-3{\bar r}_{\text{crit}}+2{\bar Q}^2}
\over{{\bar r}^3_{\text{crit}}-{\bar Q}^2{\bar r}^2_{\text{crit}}}}}\ .
\end{equation}
Substituting (\ref{Eq44}) into Eq. (\ref{Eq43}), one obtains the cubic equation
\begin{equation}\label{Eq45}
{\bar r}^3-6{\bar r}^2+9{\bar Q}^2{\bar r}-4{\bar Q}^4=0\ \ \ \ \text{for}\ \ \ \ {\bar r}={\bar r}_{\text{crit}}\  ,
\end{equation}
whose real solution can be expressed in the compact mathematical form
\begin{equation}\label{Eq46}
{\bar r}_{\text{crit}}=2+{\cal F}({\bar Q})+(4-3{\bar Q}^2)\cdot{\cal F}^{-1}({\bar Q})\  ,
\end{equation}
where
\begin{equation}\label{Eq47}
{\cal F}({\bar Q})\equiv\sqrt[3]{2{\bar Q}^4-9{\bar Q}^2+{\bar Q}^2\sqrt{4{\bar Q}^4-9{\bar Q}^2+5}+8}\  .
\end{equation}

The explicit functional dependence of the black-hole-field critical mass 
parameter ${\bar\mu}_{\text{crit}}({\bar Q})$ 
on the dimensionless charge parameter ${\bar Q}$ of the central black hole can be obtained directly 
by substituting the analytically derived relation (\ref{Eq46}) into Eq. (\ref{Eq44}). 
In Table \ref{Table1} we present the values of the black-hole-field critical mass 
parameter ${\bar\mu}_{\text{crit}}({\bar Q})$ 
for various values of the dimensionless charge parameter ${\bar Q}$. 
One finds that ${\bar\mu}_{\text{crit}}({\bar Q})$ is a 
monotonically increasing function of the black-hole charge parameter ${\bar Q}$. 
As a consistency check we note that the analytically derived resonance formula (\ref{Eq44}) correctly 
yields the relation (\ref{Eq33}) in the Schwarzschild (${\bar Q}\to0$) limit and correctly 
yields the relation (\ref{Eq39}) in the extremal (maximally charged, ${\bar Q}\to1$) limit.  

\begin{table}[htbp]
\centering
\begin{tabular}{|c|c|c|c|c|c|c|}
\hline \ \ ${\bar Q}$\ \ &\ \
$0$\ \ \ &\ \ $0.2$\ \ \ &\ \ $0.4$\ \ \ &\ \ $0.6$\ \ \ &\ \ $0.8$\ \ \ &\ \ $1$\ \ \\
\hline \ \ ${\bar\mu}_{\text{crit}}\equiv {{M\mu}\over{\sqrt{l(l+1)}}}$\ \ & \ ${{1}\over{\sqrt{12}}}$
\ \ & \ $0.2903$\ \ & \
$0.2954$\ \ & \ $0.3051$\ \ & \ $0.3218$\ \ & \ ${{1}\over{\sqrt{8}}}$\ \ \\
\hline
\end{tabular}
\caption{Quasi-resonant modes of massive scalar fields in the charged Reissner-Nordstr\"om 
black-hole spacetime. We present, for various values of the black-hole charge parameter ${\bar Q}\equiv Q/M$, 
the critical values of the dimensionless mass 
parameter ${\bar\mu}\equiv {{M\mu}/{\sqrt{l(l+1)}}}$ as 
obtained directly from the {\it analytically} derived relations (\ref{Eq44}) and (\ref{Eq46}). 
The critical mass parameter ${\bar\mu}_{\text{crit}}({\bar Q})$ characterizes composed 
black-hole-field resonances which, in the double asymptotic regime $M\mu\gg1$ 
with $l\gg1$, become arbitrarily long-lived . The data presented reveals the 
fact that ${\bar\mu}_{\text{crit}}({\bar Q})$ is a 
monotonically increasing function of the black-hole dimensionless charge parameter ${\bar Q}$. } \label{Table1}
\end{table} 

\section{The regime of validity of the WKB approximation}

It is important to emphasize that the analytically derived results of the present paper for the characteristic 
resonant spectra of composed Reissner-Nordstr\"om-black-hole-massive-scalar-field systems 
were derived under the assumption that higher-order
correction terms that appear in the WKB approximation (\ref{Eq20}) can be neglected in the 
double asymptotic regime $M\mu\gg1$ with $l\gg1$ [see Eq. (\ref{Eq18})]. 

In particular, as explicitly shown
in \cite{WKB1,WKB2,WKB3,Will}, an extension of the WKB approximation
to include higher-order spatial derivatives of the composed black-hole-field radial potential yields the 
correction term
\begin{equation}\label{Eq48}
\Lambda={{1+(2n+1)^2}\over{32}}\cdot{{V^{(4)}_0}\over{V^{(2)}_0}}-
{{28+60(2n+1)^2}\over{1152}}\cdot\Big({{V^{(3)}_0}\over{V^{(2)}_0}}\Big)^2\
\end{equation}
on the right-hand-side of the resonance equation (\ref{Eq20}). 
The mathematically compact WKB resonance condition (\ref{Eq20}) that we have used in our 
analytical study of the composed Reissner-Nordstr\"om-black-hole-massive-scalar-field system 
is therefore valid provided
\begin{equation}\label{Eq49}
\Lambda\ll V_0\  .
\end{equation}
Taking cognizance of Eqs. (\ref{Eq14}), (\ref{Eq19}), and (\ref{Eq48}), 
one finds the dimensionless relation
\begin{equation}\label{Eq50}
M^2\Lambda=O[({\bar\mu}_{\text{crit}}-\bar\mu)^{-1}]\  ,
\end{equation}
which implies that the WKB resonance condition (\ref{Eq20}) 
is valid in the large-mass-large-angular-momentum regime [see Eq. (\ref{Eq19})] \cite{Notelll}
\begin{equation}\label{Eq51}
l(l+1),(M\mu)^2\gg({\bar\mu}_{\text{crit}}-\bar\mu)^{-1}\  .
\end{equation}

\section{Summary and Discussion}

We have stuided, using {\it analytical} techniques, the physical and mathematical properties of the composed 
Reissner-Nordstr\"om-black-hole-massive-scalar-field system in the dimensionless  
large-mass-large-harmonic-indices regime $M\mu\gg1$ with $l\gg1$ [see Eq. (\ref{Eq18})]. 

The main analytical results derived in this paper and their physical implications are as follows:

(1) We have proved that the characteristic timescale associated with the
relaxation dynamics of the composed Schwarzschild-black-hole-massive-scalar-field system  
in the large-mass regime (\ref{Eq18}) can be determined analytically. 
In particular, the relaxation timescale can be expressed in terms of the dimensionless mass 
parameter ${\bar\mu}\equiv {{M\mu}/{\sqrt{l(l+1)}}}$ [see Eqs. (\ref{Eq1}), (\ref{Eq27}), 
and (\ref{Eq30})] in the compact mathematical form \cite{Notenn01}:
\begin{equation}\label{Eq52}
{{\tau^{\text{Sch}}_{\text{relax}}}\over{M}}=
{{\big(1-\sqrt{1-12{\bar\mu}^2}\big)^2}\over
{\sqrt{2}{\bar\mu}^3\sqrt{\sqrt{1-12{\bar\mu}^2}-(1-12{\bar\mu}^2)}}}\  .
\end{equation}
We have exmpahszied the physically interesting fact that the relaxation timescale (\ref{Eq52}) 
becomes extremely long in the critical mass limit ${\bar\mu}\to{\bar\mu}^{\text{Sch}}_{\text{crit}}=1/\sqrt{12}$. 

(2) We have proved that the characteristic timescale associated with the
relaxation dynamics of the composed extremal-Reissner-Nordstr\"om-black-hole-massive-scalar-field system  
in the regime (\ref{Eq18}) is given by the mass-dependent dimensionless functional relation 
[see Eqs. (\ref{Eq1}), (\ref{Eq27}), and (\ref{Eq37})] \cite{Notenn02}:
\begin{equation}\label{Eq53}
{{\tau^{\text{eRN}}_{\text{relax}}}\over{M}}=
{{\big(1-\sqrt{1-8{\bar\mu}^2}\big)^3}\over
{2{\bar\mu}^3\sqrt{[1-9{\bar\mu}^2+8{\bar\mu}^4-(1-5{\bar\mu}^2)
\sqrt{1-8{\bar\mu}^2}][1-\sqrt{1-8{\bar\mu}^2}-2{\bar\mu}^2]}}}\  .
\end{equation}
Interestingly, from Eq. (\ref{Eq53}) one deduces that the relaxation timescale 
of the composed extremal-black-hole-scalar-field system becomes extremely long in the critical mass limit 
${\bar\mu}\to^{\text{eRN}}_{\text{crit}}=1/\sqrt{8}$. 

(3) It has been shown that, for a given value ${\bar Q}\equiv Q/M$ of the dimensionless charge parameter 
of the central Reissner-Nordstr\"om black hole, the composed 
black-hole-field system is characterized by the existence of quasi-resonant modes with extremely long lifetimes. 
In particular, we have proved that the dimensionless spectrum $\{M\mu_{\text{crit}}({\bar Q},l)\}$ of 
critical black-hole-field masses that characterize the long-lived resonances of the system 
can be determined analytically [see Eqs. (\ref{Eq27}), 
(\ref{Eq44}), (\ref{Eq46}), and (\ref{Eq47})],
\begin{equation}\label{Eq54}
{M\mu}_{\text{crit}}({\bar Q})=\sqrt{l(l+1)}\cdot\sqrt{{{{\bar r}^2_{\text{crit}}-3{\bar r}_{\text{crit}}+2{\bar Q}^2}
\over{{\bar r}^3_{\text{crit}}-{\bar Q}^2{\bar r}^2_{\text{crit}}}}}\  ,
\end{equation}
in the dimensionless large-mass-large-harmonic-indices regime (\ref{Eq18}), where
\begin{equation}\label{Eq55}
{\bar r}_{\text{crit}}({\bar Q})=2+\sqrt[3]{2{\bar Q}^4-9{\bar Q}^2+{\bar Q}^2\sqrt{4{\bar Q}^4-9{\bar Q}^2+5}+8}+(4-3{\bar Q}^2)
\cdot\Bigg[\sqrt[3]{2{\bar Q}^4-9{\bar Q}^2+{\bar Q}^2\sqrt{4{\bar Q}^4-9{\bar Q}^2+5}+8}\Bigg]^{-1}\  .
\end{equation}

It is important to emphasize that the physically interesting {\it numerical} studies \cite{ELL1,ELL2} have shown that 
long-lived quasi-resonances exist in composed black-hole-massive-field systems. 
However, to the best of our knowledge, the functional relation (\ref{Eq54}) derived in the present compact paper 
provides the first {\it analytical} formula 
for the critical mass spectrum $\{M\mu_{\text{crit}}({\bar Q},l)\}$ 
that characterizes the long-lived resonant modes of the 
composed Reissner-Nordstr\"om-black-hole-massive-scalar-field system.

(4) Finally, we would like to stress the fact that the analytically derived results presented in this paper imply 
that, although static matter configurations 
which are made of minimally coupled scalar fields cannot be supported 
by Reissner-Nordstr\"om black holes \cite{Whee,Car,Bek1,Chas,BekVec,Hart,Hod11}, the 
composed Reissner-Nordstr\"om-black-hole-massive-scalar-field system can have arbitrarily long-lived 
resonant modes. 
In particular, in the dimensionless  
large-mass-large-harmonic-indices regime $M\mu\gg1$ with $l\gg1$, the long-lived black-hole-field 
quasi-resonances are characterized by the dimensionless critical mass spectrum (\ref{Eq54}) with (\ref{Eq55}).  

\bigskip
\noindent
{\bf ACKNOWLEDGMENTS}
\bigskip

This research is supported by the Carmel Science Foundation. I would
like to thank Yael Oren, Arbel M. Ongo, Ayelet B. Lata, and Alona B.
Tea for helpful discussions.



\begin{thebibliography}{99}
\bibitem{Notescc} It is physically important to emphasize the fact that some minimally-coupled 
stationary matter fields can be supported in asymptotically flat black-hole spacetimes with 
regular horizons (see \cite{Hodrc,HerR} and references therein).
\bibitem{Notetails} It is interesting to note that, in asymptotically flat spacetimes, the dynamics of 
linearized perturbation fields are also characterized by asymptotically decaying (with an 
inverse power-law functional dependence of the form 
$\psi\sim t^{-p}$ with $p>0$) wave tails, see \cite{Tails} and references therein.
\bibitem{Noteiw} Here $\Im\omega$ is the imaginary part of the complex quasinormal resonant frequency.
\bibitem{Noteunit} We shall use natural units in which $G=c=\hbar=1$.
\bibitem{Notemm} Note that the field parameter $\mu$ stands for $\mu/\hbar$ and it therefore has 
the dimensions of length$^{-1}$.
\bibitem{Notegen} One may assume $Q>0$ without loss of generality.
\bibitem{Noteom} The integer parameters $m$ and $l$ (with $l\geq |m|$) 
are respectively the azimuthal harmonic index and the spherical harmonic index of the massive 
scalar field. For brevity, we shall henceforth omit these field indices.
\bibitem{Notesy} Note that the angular scalar functions $S_{lm}(\theta)$ are the 
associated Legendre polynomials $Y_l^m(\cos\theta)$.
\bibitem{Norery} Note that the radial differential relation (\ref{Eq14}) maps the coordinate $r\in[r_+,\infty]$ 
to $y\in[-\infty,\infty]$.
\bibitem{Notesd} Here we have used the relation $d^2V/dy^2=(d^2V/dr^2)\cdot(dr/dy)^2$ at the 
location $r=r_0$ of the radial peak of the composed black-hole-field potential 
[in which case $[dV/dr]_{r=r_0}=0$, see Eqs. (\ref{Eq14}) and (\ref{Eq21})].
\bibitem{Notetr1} Note that the two dimensionless radii (\ref{Eq26}) coincide at ${\bar r}^{\text{Sch}}_{\text{crit}}=6$ 
in the critical mass limit ${\bar\mu}\to{\bar\mu}^{\text{Sch}}_{\text{crit}}={{1}\over{\sqrt{12}}}$.
\bibitem{Notetr2} Note that the two dimensionless radii (\ref{Eq35}) coincide 
at ${\bar r}^{\text{eRN}}_{\text{crit}}=4$ in the critical mass 
limit ${\bar\mu}\to{\bar\mu}^{\text{eRN}}_{\text{crit}}={{1}\over{\sqrt{8}}}$.
\bibitem{Notelll} From Eq. (\ref{Eq51}) one finds that, 
in the large-mass-large-angular-momentum regime $M\mu,l\gg1$, 
the analytically derived formula (\ref{Eq30}) can describe 
long-lived black-hole-field resonant modes with $1/(M\mu)^{1/2}\ll{\bar\omega}_{\text{I}}\ll1$.
\bibitem{Notenn01} Here we have used Eq. (\ref{Eq30}) with $n=0$ for the fundamental (least damped) 
black-hole-field resonant mode.
\bibitem{Notenn02} Here we have used Eq. (\ref{Eq37}) with $n=0$ for the fundamental (least damped) 
black-hole-field resonant mode.
\end{thebibliography}
\end{document}